# Security Evaluation of Laser-Phase-Noise Quantum Random Number Generators with Intrinsic Correlations

**Y. Gao[1]; J.H Li[2]; J.L. Liu[1]; M. Wu[1]; L.J. Liu[1]; H. Wang[1]; Y. Pan[1]; W. Huang[1]; Y. Li[1]; J. Yang[1,2]; B.J. Xu[1]**

**Abstract**

Quantum random number generators are essential for achieving information-theoretical security in modern cryptographic systems. Among various implementations, laser phase noise schemes are widely favored for their simple architecture and high integration potential. However, the intrinsic correlations in the raw data are often neglected, which violates the independent and identically distributed assumption and potentially compromises system security. In this work, we establish an analytical model of correlation and formulate an analytical expression for the conditional min-entropy in the presence of intrinsic correlations to accurately quantify the genuinely extractable randomness. The validity of our theoretical model is confirmed by numerical simulations and experimental results, exhibiting excellent agreement. Under typical setups it is shown that neglecting intrinsic correlations leads to an overestimation of extractable randomness by approximately 46%. This work provides a valuable theoretical framework for designing compact, high-performance quantum random number generators with rigorous security analysis.

**Affiliations**

1. National Key Laboratory of Security Communication, Institute of Southwestern Communication, Chengdu 610041, China.
2. State Key Laboratory of Information Photonics and Optical Communications, School of Electronic Engineering, Beijing University of Posts and Telecommunications, Beijing 100876, China.

**Corresponding authors:** J. Yang (everest1573@163.com); B.J. Xu (xbjpku@163.com)

## Introduction

Random numbers serve as a fundamental source of modern information security, playing an indispensable role in cryptography, statistical sampling, numerical simulations and quantum information processing[1–5]. Unlike traditional pseudo-random number generators (PRNGs) that rely on deterministic algorithms and possess inherent vulnerabilities[6], quantum random number generators (QRNGs) exploit the intrinsic indeterminacy of quantum mechanics to produce theoretically unpredictable entropy[3,7]. Among various implementations of QRNGs[8–29], schemes based on laser phase noise have emerged as one of the most promising candidates for high-speed and low-cost applications due to their structural simplicity and excellent compatibility with photonic integration technology[24–28].

The growing demand for digital and chip-scale integration has made the miniaturization of QRNGs an inevitable trend. However, single-laser phase noise schemes face a fundamental physical challenge when moving toward compact architectures. Traditionally, to reduce temporal correlations and ensure high-quality randomness, the delay time between two arms is required to be significantly longer than the laser coherence time, which leads to the use of extensive optical delay line that increase the device size and hinder integration[18,19]. Conversely, shortening the delay to achieve a compact design inevitably introduces significant temporal correlations into the raw data[18–20]. The intrinsic correlations, rooted in the underlying physical overlap of phase evolution, represents a primary obstacle to the development of high-performance integrated QRNGs based on laser phase noise.

The paramount metric for QRNGs rely on provable security, which is typically guaranteed through entropy evaluation for randomness quantification within an information-theoretic framework[3,16]. While the min-entropy has been generally employed as the standard methodology for entropy evaluation, its validity to date relies on the independent and identically distributed (i.i.d.) assumption of the raw data[16–19,29–32]. However, due to the intrinsic correlations of raw data in single-laser phase noise schemes, the min-entropy evaluation based on the i.i.d. assumption will overestimate the system security. To provide a rigorous security guarantee, it is imperative to employ conditional min-entropy that accounts for intrinsic correlations. Therefore, further investigation into the physical mechanisms of intrinsic correlations and the methods of randomness quantification are foundational to establishing a rigorous security evaluation framework. Despite extensive research on laser-phase-noise based QRNGs, most existing studies either neglect intrinsic correlations or lack a precise quantitative treatment[16–19,29–32], leaving an unsolved open question in the rigorous evaluation of their practical security.

In this work, we establish a rigorous theoretical model for single-laser phase noise schemes and derive the analytical expression for the correlation coefficients of the raw data. This model explicitly quantifies the dependence of correlations on crucial system parameters, which is convincingly validated via the excellent agreement between theoretical predictions and numerical simulations. Furthermore, we formulate an analytical expression for the conditional min-entropy, enabling an accurate quantification of extractable randomness even in the presence of significant intrinsic correlations. Specifically, accounting for electronic noise in typical setups, we find that neglecting intrinsic correlations results in a 46% overestimation of extractable randomness. Consequently, this research provides an valuable theoretical foundation for developing compact, high-performance QRNGs that ensures rigorous security analysis while pursuing device miniaturization.

## THEORY MODEL AND NUMERICAL SIMULATION

### Physical model

A typical experimental setup for QRNG schemes based on detecting the laser phase noise is shown in Fig. 1(a), where the random phase shift of the optical signal from a single laser at two different times is distilled via an unbalanced Mach-Zehnder interferometer. Mathematically, the electric field of the laser output is described as $E(t) = E_0\exp[i(\omega_0 t + \phi(t))]$, where $\phi(t)$ represents the phase fluctuations of the laser source and $\omega_0$ is the angular frequency. The output voltage $V(t)$, obtained from the photodetector ($PD_1$) after removing the DC components, can be expressed as [21]:

$$V(t) = A\cos[\omega_0 T_d + \Delta\phi(t, T_d)]. \quad (1)$$

Here, A is a conversion factor that includes the amplitude of the optical signal, the attenuation of the optical path and the response of the photodiode, etc. $\Delta\phi(t, T_d) = \phi(t) - \phi(t - T_d)$ denotes the random phase shift of the laser source between time $t$ and $t - T_d$, $T_d$ is the path delay between two arms. By stabilizing this static phase difference at $[2m\pi + \pi/2]$ (where $m$ is an integer), the PD output simplifies to:

$$V(t) = A\sin[\Delta\phi(t, T_d)]. \quad (2)$$

In practice, the total continuous-time voltage signal $Y(t)$ fed into the analog-to-digital converter (ADC) incorporates not only the purely quantum signal derived from phase fluctuations, but also classical electronic noise. Let $X(t) \equiv V(t)$ represent this raw data of quantum signal. As shown in previous works, under a reasonable sampling rate (e.g., $f_s \leq 2BW$, where $BW$ is the bandwidth of the quantum signal which is the output of the PD), the quantum signal $X(t)$ is generally assumed to be independent and identically distributed (i.i.d.). However, overlooking its intrinsic correlations leads to an overestimation of extractable randomness, potentially compromising the system's security.

Consequently, to rigorously quantify the system's security, we must account for the intrinsic correlations. we explicitly decompose the quantum signal into an uncorrelated component $X_{non}(t)$ and a correlated $X_c(t)$, such that $X(t) = X_{non}(t) + X_c(t)$. Combined with the classical electronic noise $E(t)$, the physical model for the total detected signal $Y(t)$ is then given by:

$$Y(t) = X(t) + E(t) = X_{non}(t) + X_c(t) + E(t). \quad (3)$$

Here, $X_{non}(t)$ is the strictly non-correlated quantum signal that contributes to the attainable min-entropy of the system, $X_c(t)$ denotes the quantum signal with intrinsic correlations, and $E(t)$ refers to the statistically independent classical noise (e.g., electronic noise from the detector and ADC).

To systematically analyze the statistical properties and intrinsic correlations of the raw data, we model the phase noise $\phi(t)$ as a Wiener process[33], characteristic of a laser with a Lorentzian spectrum. Consequently, the phase increment $\Delta\phi(t, T_d)$ is modeled as Gaussian random variable satisfying $\Delta\phi(t, T_d) = \phi(t) - \phi(t - T_d) \sim N\left(0, \frac{2T_d}{\tau_c}\right)$[34,35], where $\tau_c = \frac{1}{\pi\Delta\nu}$ represents the coherence time of the laser with a linewidth $\Delta\nu$[36].

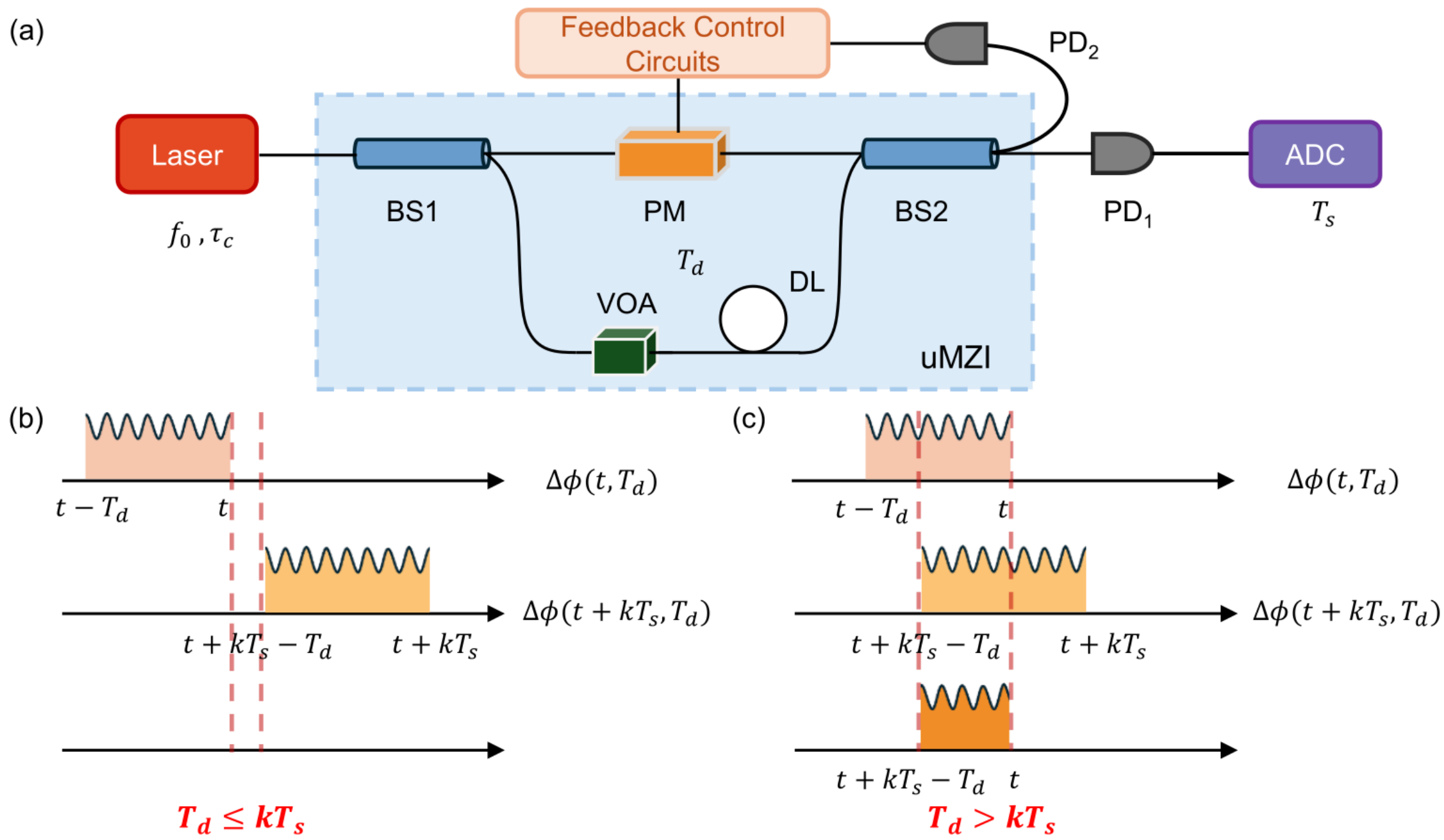

***Figure 1.*** Experimental scheme and phase evolution analysis of the single-laser QRNG. (a) The schematic diagram of the experimental setup based on detecting the phase noise from a single laser. (b) and (c) Temporal relationship between the phase shifts $\Delta\phi(t, T_d)$ and $\Delta\phi(t + kT_s, T_d)$ under the scenario of $T_d \leq kT_s$ and $T_d > kT_s$, respectively. BS: beam splitter, PM: phase modulator, PD: photo-detector, uMZI: unbalanced Mach-Zehnder interferometer, DL: delay line, VOA: variable optical attenuator, ADC: analog-to-digital convertor.

The intrinsic correlations between the $n$-th sample at $t$ and the $n + k$-th sample at $t + kT_s$ is strictly dictated by the temporal overlap of their respective phase evolution intervals, $I_1 = [t - T_d, t]$ and $I_2 = [t + kT_s - T_d, t + kT_s]$, as illustrated in Fig. 1(b) and (c). When $T_d \leq kT_s$, the intervals are disjoint, resulting in noncorrelation due to the independent increment property of the Wiener process. Conversely, for $T_d > kT_s$, the intervals share a common physical process of phase evolution over a duration $T_{overlap,k} = T_d - kT_s$, thereby inducing correlation[37]. The covariance between the phase increment $\Delta\phi(t, T_d)$ and $\Delta\phi(t + kT_s, T_d)$ is governed by:

$$\mathrm{Cov}\big(\Delta\phi(t, T_d), \Delta\phi(t + kT_s, T_d)\big) = 2\pi\Delta\nu T_{overlap,k}. \tag{4}$$

Since these increments originate from the same laser source and follow a zero-mean jointly Gaussian distribution, applying the properties of the joint characteristic function allows us to derive the covariance of the interference signals (See Appendix A for the detailed derivation):

$$\begin{aligned} &\mathrm{Cov}\big(\sin\Delta\phi(t, T_d), \sin\Delta\phi(t + kT_s, T_d)\big) \\ &= e^{-\frac{2T_d}{\tau_c}} \sinh\left(\frac{2}{\tau_c}(T_d - kT_s)\right). \end{aligned} \tag{5}$$

Finally, the correlation coefficient $\rho_X(k)$ of quantum signal is obtained by dividing the covariance above by the signal variance $\mathrm{Var}[\sin\Delta\phi(t,T_d)] = \frac{1}{2}\left(1 - e^{-\frac{4T_d}{\tau_c}}\right)$. Expressed in terms of the laser coherence time $\tau_c$, this yields the universal analytical expression:

$$\rho_X(k) = \frac{\sinh\left(\frac{2(T_d - kT_s)}{\tau_c}\right)}{\sinh\left(\frac{2T_d}{\tau_c}\right)} \tag{6}$$

Crucially, the analytical expression derived above is universally applicable regardless of the relationship between $T_d$ and $\tau_c$, as it relies on the exact statistical properties of the Wiener process rather than small-angle approximations (i.e., $\sin\Delta\phi \simeq \Delta\phi$)[20,21].

### Numerical simulation

To comprehensively validate the accuracy and universality of the proposed theoretical model, we conducted numerical simulations using typical experimental parameters to faithfully reproduce the random number generation process based on laser phase noise. Subsequently, we performed a rigorous statistical analysis on the large-scale raw data generated by the simulation. Specifically, we calculated the $k$-th order autocorrelation coefficients and compared these simulation results directly with the theoretical values predicted by the analytical formula (Eq. 6), as illustrated in Fig. 2, which details the comparison results under three distinct physical regimes. Figs. 2 (a) and (d) display the $k$-th order autocorrelation coefficients for the regime where $T_d \gg \tau_c$ ($T_d = 1\,\mu$s, $\tau_c = 100$ ns, $T_s = 50$ ns), using sequence lengths of $10^8$ and $10^{12}$, respectively. Figs.2 (b) and (e) correspond to the $T_d \simeq \tau_c$ ($T_d = 300$ ns, $\tau_c = 300$ ns, $T_s = 15$ ns), while Figs.2 (c) and (f) represent the $T_d \ll \tau_c$ ($T_d = 1$ ns, $\tau_c = 300$ ns, $T_s = 0.05$ ns). In all plots, red stars represent theoretical predictions, while blue triangles denote numerical simulation results.

From the figure, it is clearly observable that the autocorrelation coefficients derived from numerical simulations align closely with the theoretical prediction curves, regardless of the parameter conditions. This agreement strongly demonstrates the excellent universality of the analytical model in accurately characterizing the system's correlation properties across various operating scenarios.

It is worth noting that in the case of $T_d \gg \tau_c$ as shown in Figs. 2 (a) and (d), a slight deviation between the simulated values and the theoretical predictions becomes apparent when the lag $k$ is large. This discrepancy does not indicate a deficiency in the analytical model but rather stems from insufficient statistical precision limited by the finite data length. By comparing the simulation results obtained with data lengths of $10^8$ and $10^{12}$, it is evident that the results converge significantly toward the theoretical predictions as the data length increases. This confirms that the observed deviations at large lags are primarily due to statistical fluctuations inherent in finite datasets, which are systematically suppressed as the total number of samples grows, rather than any limitation of the theoretical framework[38–41].

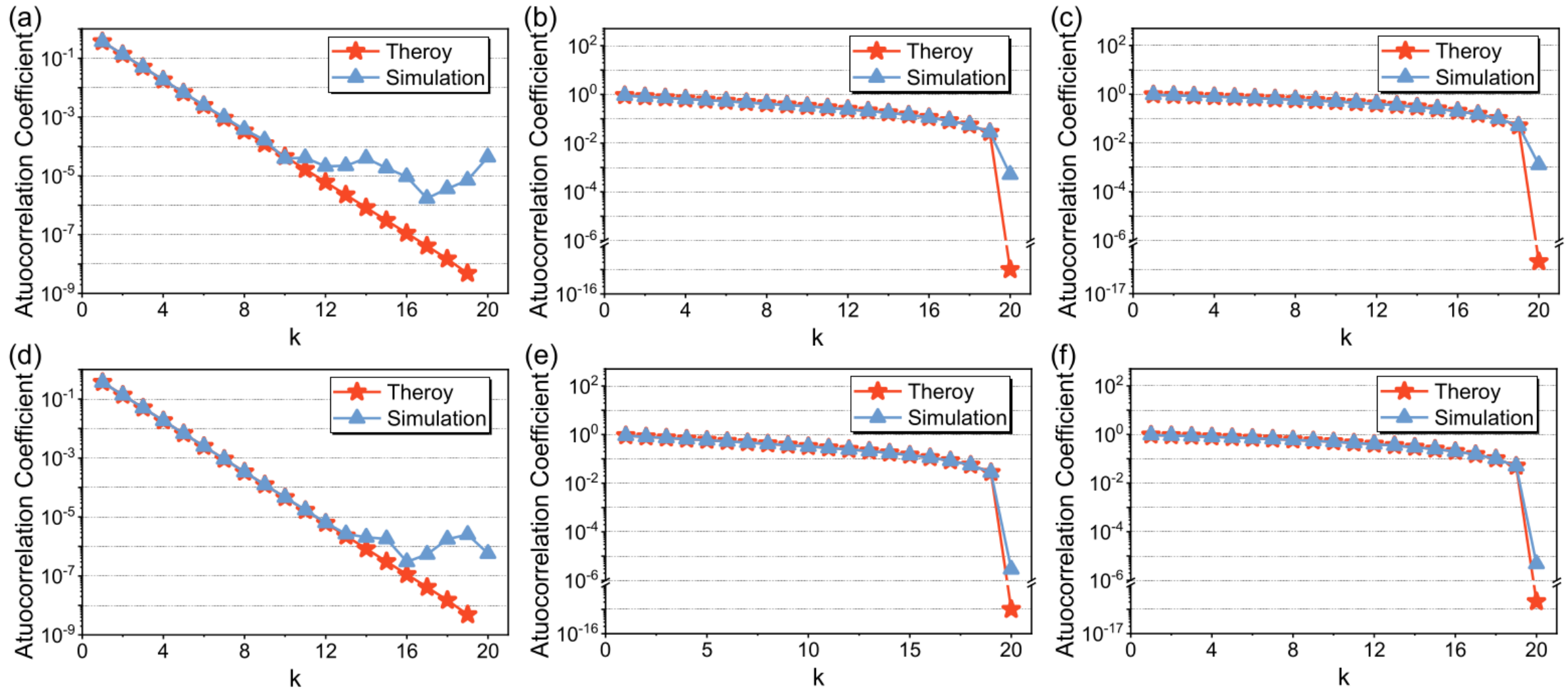

**Figure 2.** Comparison of the theoretical and simulated k-th order autocorrelation coefficients under different parameter conditions. (a), (d) Results for $T_d \gg \tau_c$ with data lengths of $10^8$ and $10^{12}$, respectively. (b), (e) Results for $T_d \simeq \tau_c$ with data lengths of $10^8$ and $10^{12}$ respectively. (c), (f) Results for $T_d \ll \tau_c$ with data lengths of $10^8$ and $10^{12}$, respectively. Red stars and blue triangles represent the theoretical predictions and numerical simulation results, respectively.

## Entropy evaluation

**Accounting for correlation.** The preceding correlation analysis confirms that, under conditions of temporal overlap, the raw data generated by the system exhibit non-negligible correlations, which indicates a violation of the i.i.d. assumption required for ideal generation of random numbers. Consequently, employing min-entropy for randomness quantification is inappropriate and may lead to security vulnerabilities. To ensure information-theoretic security, we adopt conditional min-entropy, $H_{\min}(X_m^{\text{dis}} \mid \boldsymbol{X}_t^{\text{dis}})$, which quantifies the uncertainty of a target variable conditioned on its correlated neighbors $\boldsymbol{X}_t$. Since the raw data originates from laser phase noise, which follows a Gaussian distribution, the underlying continuous random sequence $X = \{X_1, X_2, \dots\dots, X_n\}$ is rigorously modeled as a multivariate Gaussian process fully described by the covariance matrix $\Sigma$.

Factually, the continuous signal from the photodetector is quantized into discrete values by an ADC with a step size $\Delta_X$. To model this process, the discrete conditional probability $P(X_m^{\text{dis}} = x_m \mid \boldsymbol{X}_t^{\text{dis}} = \boldsymbol{x}_t)$ is obtained by integrating the continuous conditional PDF, $f(X_m \mid \boldsymbol{X}_t)$, over the quantization bins $D_m$ and $D_t$ associated with the $m$-th sample (See Appendix B for the detailed derivation):

$$
\begin{aligned}
P(X_m^{\text{dis}} \mid \boldsymbol{X}_t^{\text{dis}}) &= \frac{P(X_m^{\text{dis}} = x_m, \boldsymbol{X}_t^{\text{dis}} = \boldsymbol{x}_t)}{P(\boldsymbol{X}_t^{\text{dis}} = \boldsymbol{x}_t)} \\
&= \frac{\int_{D_m} \int_{D_t} f(\boldsymbol{X}_{\text{joint}})\, dX_m\, d\boldsymbol{X}_t}{\int_{D_t} f\,(\boldsymbol{X}_t)\, d\boldsymbol{X}_t}.
\end{aligned} \tag{7}
$$

The final expression for the conditional min-entropy is given by minimizing the uncertainty (or maximizing the prediction probability) over all possible outcomes:

$$
H_{\min}(X_m^{\text{dis}} \mid \boldsymbol{X}_t^{\text{dis}}) = -\log_2 \left( \max_{x_m, \boldsymbol{x}_t} P(X_m^{\text{dis}} \mid \boldsymbol{X}_t^{\text{dis}}) \right). \tag{8}
$$

The analysis presented above considers the ideal scenario where electrical noise is negligible. However, in practical experimental settings, electrical noise must be incorporated into the evaluation. A detailed procedure for calculating conditional min-entropy in the presence of intrinsic correlations and electrical noise is provided in the **Practical consideration**.

Furthermore, while Eq. (8) provides a rigorous framework for evaluating conditional min-entropy via multi-dimensional integration, its numerical implementation often entails significant computational overhead and lacks a direct intuitive link to the underlying physical parameters. To derive a more tractable analytical expression, we define the correlation range $\beta = \left\lceil \frac{T_d}{T_s} \right\rceil$, which represents the maximum number of historical or future samples correlated with $X_m$. The correlation coefficient $\rho_X(k)$ is a monotonically decreasing function of the $k$ as an increase in the sampling interval leads to a reduction in the temporal overlap $T_{overlap,k}$. Consequently, the uncertainty of $X_m$ is primarily constrained by its most strongly correlated neighbor $X_{m-k}$ ($k$=1). Leveraging the Markovian property of the source, we can reduce the complex $2\beta$-dimensional conditional model into a bivariate correlation framework[30,42].

Under this simplified bivariate framework, we further assume an $n$-bit ADC with a scale range of $R_X = A$ and $\Delta_X = \frac{2A}{2^n}$. When $\Delta_X$ is significantly smaller than the signal's standard deviation, the maximum discrete conditional probability is approximated by the peak value of the continuous conditional PDF[43], i.e., $P_{\max}(X_m^{\text{dis}} = x_m \mid X_{m-k}^{\text{dis}} = x_{m-k}) \simeq f(X_m = \mu | X_{m-k}) \cdot \Delta_X$. By exploiting the peak characteristics of the Gaussian PDF, we derive the conditional min-entropy in terms of the physical parameters:

$$H_{\min}\left(X_m^{\text{dis}} \mid X_{m-k}^{\text{dis}}\right) = \log_2\left(\frac{A\sqrt{\pi\left(1 - e^{-\frac{4T_d}{\tau_c}}\right)\left(1 - \rho_X^2(k)\right)}}{\Delta_X}\right) \tag{9}$$

This simplified model explicitly elucidates the competing effects among the delay time $T_d$, the laser coherence time $\tau_c$, and the sampling interval $kT_s$, providing a theoretical foundation for optimizing the trade-off between system miniaturization and high-speed generation in practical QRNG implementations. Under the i.i.d. assumption with the raw data, adjacent sampling points are considered statistically independent, causing the conditional PDF to reduce to the marginal distribution $f(X_m)$. Thus, the min-entropy can be expressed as:

$$H_{\min}^{\text{iid}} = \log_2\left(\frac{A\sqrt{\pi(1 - e^{-\frac{4T_d}{\tau_c}})}}{\Delta_X}\right). \tag{10}$$

**Practical consideration.** The preceding analysis, predicated on the ideal condition of negligible electrical noise, provides a clear theoretical foundation for understanding the physical mechanisms underlying QRNGs. However, in practical experimental setup, photodetectors and associated electronic components inevitably introduce classical electrical noise. From a rigorous cryptographic security perspective, such classical noise is typically regarded as untrusted side information that could be fully accessible to or predictable by a potential eavesdropper. If the evaluation methods derived for the ideal scenario are directly applied to the actual measured signals containing electrical noise, the system would erroneously attribute classical noise fluctuations to genuine quantum randomness. This would severely overestimate the amount of extractable secure randomness, thereby introducing critical security vulnerabilities.

Therefore, to provide rigorous security guarantees in practical applications, it is imperative to incorporate electrical noise into the entropy evaluation model (the detailed theoretical analysis is provided in the Appendix C). Given complete knowledge of the electrical noise instance $E = e$, the conditional variance of the measured signal $Y$ reduces to the variance of quantum signal $X$. However, in a practical experimental setting, the quantum variance $\sigma_X^2$ and its intrinsic correlations $\rho_X(k)$ are not directly observable. Instead, we can only experimentally determine the statistical parameters of the total digitized signal $Y$ and independently characterize the electrical noise variance $\sigma_E^2$.

Let $\sigma_Y^2$ be the measured variance of the total signal $Y$, and $\rho_Y(k)$ be the measured correlation coefficient at the $k$. The electrical noise $E$ is modeled as Gaussian random process with a known variance $\sigma_E^2$. Because the quantum signal $X$ and the electrical noise $E$ are physically independent, the variance of the total signal is the linear sum of their variances. Thus, the unobservable quantum variance can be extracted as[28]:

$$\sigma_Y^2 = \sigma_X^2 + \sigma_E^2. \tag{11}$$

Moreover, since the electrical noise exhibits no correlation between different temporal samples, the covariance of the total signal is contributed entirely by the quantum signal:

$$\begin{aligned} \mathrm{Cov}(Y_m, Y_{m-k}) &= \mathrm{Cov}(X_m + E_m, X_{m-k} + E_{m-k}) \\ &= \mathrm{Cov}(X_m, X_{m-k}), \end{aligned} \tag{12}$$

$$\rho_Y(k)\sigma_Y^2 = \rho_X(k)\sigma_X^2. \tag{13}$$

The peak of the conditional Gaussian PDF is given by $1/\sqrt{2\pi\sigma_{Y_{cond}}^2}$, where $\sigma_{Y_{cond}}^2$ is the conditional variance of the quantum signal. Using the bivariate model, this conditional variance is $\sigma_{Y_{cond}}^2 = \sigma_X^2\left(1-\rho_X^2(k)\right)$[44]. By expanding this term and substituting the measurable parameters from Eq. (11 - 12), we obtain the conditional variance explicitly parameterized by the electrical noise, intrinsic correlations and total signal:

$$\sigma_{Y_{cond}}^2 = \sigma_X^2 - \frac{(\rho_X(k)\sigma_X^2)^2}{\sigma_X^2} = (\sigma_Y^2 - \sigma_E^2) - \frac{(\rho_Y(k)\sigma_Y^2)^2}{\sigma_Y^2 - \sigma_E^2}. \tag{14}$$

Following the same simplification logic used for the ideal case, we focus on the scenario involving nearest-neighbor correlation to achieve a precise and tractable evaluation (See Appendix C for the detailed derivation). By quantifying the genuine quantum uncertainty under the assumption that an eavesdropper possesses full knowledge of the classical electrical noise. The ADC quantization step size $\Delta_Y$ must be explicitly parameterized by the system's physical and engineering constraints. In practical QRNG implementations, the ADC full-scale range $R_Y$ is typically configured to encompass the stochastic fluctuations of the total measured signal $Y$ to prevent signal clipping. By defining the power signal-to-noise ratio as $\mathrm{SNR} = \frac{\sigma_X^2}{\sigma_E^2}$ and adopting a conservative range of $R_Y = \pm 6\sigma_Y$, the quantization step for an $n$-bit ADC is given by $\Delta_Y = \frac{12\sigma_Y}{2^n}$, we obtain

$$\Delta_Y = \frac{12A}{2^n}\sqrt{\frac{\left(1-e^{-\frac{4T_d}{\tau_c}}\right)(1+\mathrm{SNR})}{2\mathrm{SNR}}}. \tag{15}$$

Finally, by substituting this noise-incorporated conditional variance $\sigma_{Y_{cond}}^2$ into the approximated min-entropy definition, $H_{\min} \simeq -\log_2(f_{peak}\cdot\Delta_Y)$, we arrive at the simplified analytical expression for the conditional min-entropy:

$$\begin{aligned} &H_{\min}(Y_m^{\mathrm{dis}} \mid Y_{m-k}^{\mathrm{dis}}, E) \\ &= \log_2\left(\frac{A\sqrt{\pi\left(1-e^{-\frac{4T_d}{\tau_c}}\right)\left(1-\rho_X^2(k)\right)}}{\Delta_Y}\right). \end{aligned} \tag{16}$$

This simplified model explicitly elucidates the penalizing effect of the classical electrical noise on the system's randomness. It reveals that the extractable secure randomness must be strictly evaluated by discounting the electrical noise variance $\sigma_E^2$ from the total measured signal, providing a rigorous and direct metric for optimizing practical QRNG implementations.

## EXPERIMENTAL DEMONSTRATION

### Experiment setup

The experimental validation of the proposed theoretical model was performed using the setup illustrated in Fig. 1(a). Specifically, a 1550 nm single-mode continuous wave (CW) distributed-feedback (DFB) diode laser serves as the entropy source, exhibiting a linewidth of 6 MHz, which corresponds to a coherence time of 53.05 ns. The emitted light is attenuated by an optical attenuator and coupled into an uMZI constructed using two symmetric 50:50 fiber couplers. To mitigate phase shift introduced during the experiment, a PM is employed in one arm for active phase stabilization, ensuring the system remains locked at the quadrature point ($\omega_0 T_d = 2m\pi + \pi/2$). In the other arm, a VOA and DL are employed to precisely set power intensity and the time delay.

Subsequently, the optical signals from the two paths interfere at the second coupler partitioned into two paths. In the generation path, the signal is detected by $PD_1$ (Thorlabs PDB480C-AC) with a bandwidth of 1.6 GHz and subsequently sampled at 2 GSamples/s by an 8-bit ADC integrated within an oscilloscope (Keysight DSOV084A, 8 GHz bandwidth) to generate the raw data. In contrast, the other portion is captured by $PD_2$ and utilized as a feedback signal to a PM, stabilizing the phase shift between the two arms of the uMZI. Finally, to explicitly address the demand for miniaturization in practical applications and to test the model under strongly correlated conditions, a delay time significantly shorter than the coherence time was intentionally selected. Consequently, the delay time was set to $T_d = 5.25$ ns, corresponding to a path length difference of approximately 1.12 m.

### Experimental results and discussion

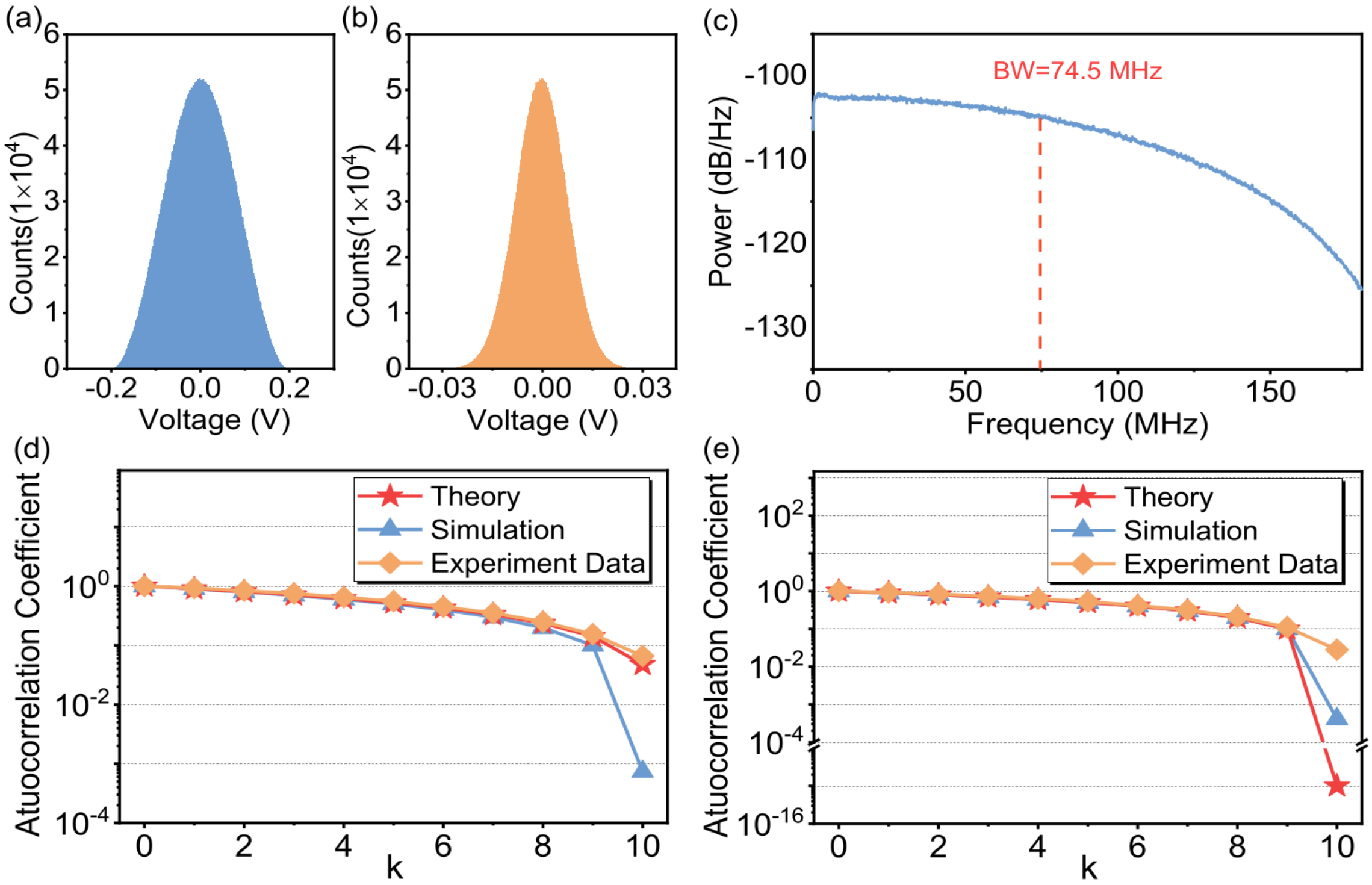

**Figure 3.** The experimental results of the QRNG based on a laser phase noise. (a) and (b) Histograms of the raw data (blue bar) and the electrical noise (orange bar), which follow a quasi-Gaussian and a Gaussian distribution, respectively. (c) Power spectrum density of the generated raw data, showing a 3-dB bandwidth of approximately 74.5 MHz. (d) and (e) Autocorrelation coefficients under the conditions of $T_d \neq kT_s$ ($T_d = 5.25\ ns$) and $T_d = kT_s$ ($T_d = 5\ ns$), respectively. Red stars, blue triangles, and orange diamonds represent the theoretical predictions, numerical simulations, and experimental data, respectively.

The experimental results are illustrated in Fig. 3, where Fig. 3(a) and (b) present the histograms of the raw data and the electrical noise, following a quasi-Gaussian and a Gaussian distribution, respectively. The frequency domain characteristics are illustrated in Fig. 3(c). The power spectrum density of the detection results, obtained with a fixed laser coherence time, exhibits a measured 3-dB bandwidth of approximately 74.5 MHz. To rigorously verify the accuracy of our proposed unified correlation model, we calculated the $k$-th order autocorrelation coefficients of the raw data and compared the experimental results with both theoretical predictions and numerical simulations, as shown in Fig. 3(d). In the plots, red stars, blue triangles, and orange diamonds represent the theoretical predictions, numerical simulations, and experimental data, respectively. It is clearly observed that the experimental data exhibit excellent agreement with both the theoretical curve and simulation results. This consistency not only confirms the validity of our analytical expression but also demonstrates that the intrinsic correlations within the raw data are fully predictable and can be rigorously accounted for by our theoretical framework. It is worth noting that at the maximum $k$, the theoretical autocorrelation coefficient does not vanish to zero. This behavior is primarily attributed to the fact that the delay time $T_d$ is not an integer multiple of the sampling period $T_s$ ($T_d \neq kT_s$). This non-vanishing correlation arises physically from the residual temporal overlap that persists even at the maximum lag, resulting in a non-zero correlation.

Based on these findings, we adjusted the parameters to $T_d$=5 ns and $T_s$=0.5 ns, as shown in Fig. 3(e). As expected, the theoretical predictions maintain excellent agreement with the experimental data. Notably, at the maximum lag $k$, where the temporal overlap theoretically ceases, the theoretical correlation coefficient approaches zero, which is consistent with our analysis of Fig. 3(d). Furthermore, a deviation between the experimental data and theoretical predictions is observed at this maximum $k$. This discrepancy is attributed to statistical fluctuations arising from the finite data length, an observation that is fully consistent with the error analysis presented in Fig. 2.

By substituting the experimental parameters ($T_d = 5.25$ ns, $\tau_c = 53.05$ ns, $T_s = 0.5$ ns) into the analytical model, we calculate the min-entropy under the conventional i.i.d. assumption to be $H_{\min}^{\text{iid}} = 7.02$ bits/sample. However, after accounting for the intrinsic correlations of the miniaturized device, $H_{\min}(X_m^{\text{dis}} \mid X_{m-k}^{\text{dis}}) = 5.80$ bits/sample. This result indicates that relying on the i.i.d. assumption would overestimate the extractable randomness per sample by approximately 21%. Furthermore, in the presence of electrical noise, we assume a SNR of 20 dB, the $H_{\min}^{\text{iid-E}} = 6.60$ bits/sample with i.i.d. assumption[32], whereas $H_{\min}(Y_m^{\text{dis}} \mid Y_{m-k}^{\text{dis}}, E) = 4.51$ bits/sample with intrinsic correlations. In this realistic scenario, the discrepancy becomes even more pronounced, with the i.i.d. assumption overestimating the extractable randomness by approximately 46%, representing a significantly larger security gap. The results indicate that relying on the i.i.d. assumption would fundamentally overestimate the extractable randomness, thereby introducing potential security. By adopting conditional min-entropy for randomness extraction, we ensure that the generated random numbers are strictly based on the unpredictability of the quantum source, effectively reducing security risks associated with intrinsic correlations.

## CONCLUSION

In conclusion, this work establishes a rigorous theoretical model for the entropy evaluation of raw data in laser-phase-noise QRNGs. By constructing a correlation analysis model, we derive the analytical expressions for correlation coefficients and conditional min-entropy in the presence of intrinsic correlations. Furthermore, we demonstrate that under a typical setup accounting for electronic noise, applying the conventional i.i.d. assumption for entropy evaluation overestimates extractable randomness by approximately 46%, posing a serious security vulnerability. By analyzing the inherent conflict between device miniaturization and rigorous security, this research provides a valuable theoretical foundation for the development of high-performance, chip-scale QRNGs. The versatility of this model allows it to be extended to various other laser-phase-noise architectures, paving the way for the realization of more compact and certified quantum entropy modules essential for the future of miniaturized quantum secure communications.

## Acknowledgments

## Data Availability

The data that support the results of this work are available from the corresponding author on reasonable request.

## Appendices

### Appendix A. Detailed derivations of correlation coefficients

The following derivation establishes a rigorous mathematical foundation for the correlation model by detailing calculation of the correlation coefficients. Under the condition of a Wiener process, the phase drift variance accumulates linearly over time as $\mathrm{Var}\big(\phi(t+\tau)-\phi(t)\big)=2\pi\Delta\nu\tau$, which leads directly to the identical phase variances at any sampling instant, $\mathrm{Var}\big(\Delta\phi(t,T_d)\big)=\mathrm{Var}\big(\Delta\phi(t+kT_s,T_d)\big)=\sigma_X^2=2\pi\Delta\nu T_d$. When these intervals exhibit a temporal overlap, the covariance between the phase increments is determined by the duration of the shared physical process, denoted as $C_k=\mathrm{Cov}\big(\Delta\phi(t,T_d),\Delta\phi(t+kT_s,T_d)\big)=2\pi\Delta\nu T_{overlap,k}$.

Since both phase differences originate from the same laser source, they are modeled as zero-mean jointly Gaussian variables. To analyze the correlation of the resulting interference signals, we first consider the covariance as:

$$\begin{aligned}&\mathrm{Cov}\big(\sin\Delta\phi(t,T_d),\sin\Delta\phi(t+kT_s,T_d)\big)\\&=E\big[\sin\big(\Delta\phi(t,T_d)\big)\sin\big(\Delta\phi(t+kT_s,T_d)\big)\big]-E\big[\sin\big(\Delta\phi(t,T_d)\big)\big]E\big[\sin\big(\Delta\phi(t+kT_s,T_d)\big)\big].\end{aligned}\quad (A1)$$

Given that the phase increment $\Delta\phi$ follows a Gaussian distribution centered at zero, the probability density function is an even function, while the sine function is odd, resulting in a vanishing mean $E[\sin(\Delta\phi)]=0$. Consequently, the covariance simplifies to the expectation of the product, $E\big[\sin\big(\Delta\phi(t,T_d)\big)\sin\big(\Delta\phi(t+kT_s,T_d)\big)\big]$. We utilize the joint characteristic function for Gaussian variables and Euler's formula, the product expectation can be expanded as:

$$\begin{aligned}&E\big[\sin\big(\Delta\phi(t,T_d)\big)\sin\big(\Delta\phi(t+kT_s,T_d)\big)\big]\\&=\sinh\Big(\mathrm{Cov}\big(\Delta\phi(t,T_d),\Delta\phi(t+kT_s,T_d)\big)\Big)\times e^{-\frac{\mathrm{Var}\left(\Delta\phi(t,T_d)\right)+\mathrm{Var}\left(\Delta\phi(t+kT_s,T_d)\right)}{2}}.\end{aligned}\quad (A2)$$

By substituting the variance $\sigma_X^2$ and the covariance $C_k$ into the identity, the expectation of the product of the interference signals simplifies to:

$$\begin{aligned}&\mathrm{Cov}\big(\sin\Delta\phi(t,T_d),\sin\Delta\phi(t+kT_s,T_d)\big)\\&=e^{-\frac{2T_d}{\tau_c}}\sinh\Big(\frac{2}{\tau_c}(T_d-kT_s)\Big).\end{aligned}\quad (A3)$$

To normalize this result, the variance of the sampled signal must be determined. Following the zero-mean property established above, the variance simplifies to $\mathrm{Var}[\sin\Delta\phi]=E[\sin^2\Delta\phi]$. Utilizing the trigonometric identity $\sin^2\theta=[1-\cos(2\theta)]/2$, the expectation is calculated as:

$$\begin{aligned}\mathrm{Var}(\sin\Delta\phi)&=\frac{1}{2}-\frac{1}{2}E[\cos(2\Delta\phi)]\\&=\frac{1}{2}\big(1-e^{-2\sigma_X^2}\big).\end{aligned}\quad (A4)$$

Finally, the normalized autocorrelation coefficient $\rho_X(k)$ is obtained by dividing the covariance by the variance:

$$\rho_X(k) = \frac{\mathrm{Cov}\big(\sin\Delta\phi(t,T_d),\sin\Delta\phi(t+kT_s,T_d)\big)}{\sqrt{\mathrm{Var}\big(\sin\Delta\phi(t,T_d)\big)}\sqrt{\mathrm{Var}\big(\sin\Delta\phi(t+kT_s,T_d)\big)}} = \frac{\sinh\left(\frac{2(T_d-kT_s)}{\tau_c}\right)}{\sinh\left(\frac{2T_d}{\tau_c}\right)} \qquad (A5)$$

**Appendix B. conditional min-entropy in ideal scenarios**

To establish a rigorous framework for quantifying randomness in the presence of temporal correlations, we detail the derivation of the conditional min-entropy for the ideal case where electrical noise is negligible. The continuous output raw data $X = \{X_1, X_2, \dots\dots, X_n\}$ generated from the laser phase noise is rigorously modeled as a zero-mean multivariate Gaussian process. The statistical properties of this process are fully encapsulated by the covariance matrix $\Sigma$, where each element $\sigma_{ij}$ represents the covariance between the variables $X_i$ and $X_j$. Based on the physical derivation of phase-noise correlation and the signal variance $\sigma_X^2$, elements are expressed as:

$$\sigma_{ij} = \begin{cases} e^{-\frac{2T_d}{\tau_c}}\sinh\left(\frac{2T_d}{\tau_c}\right), & i=j, \\ e^{-\frac{2T_d}{\tau_c}}\sinh\left(\frac{2(T_d-|i-j|T_s)}{\tau_c}\right), & \frac{T_d}{T_s} > |i-j|, \\ 0, & \frac{T_d}{T_s} \le |i-j|. \end{cases} \qquad (B1)$$

We define a vector of correlated neighbors $\boldsymbol{X}_t = [X_{m-\beta}, X_{m-\beta+1}, \dots, X_{m-1}, X_{m+1}, \dots, X_{m+\beta-1}, X_{m+\beta}]^T$. The joint PDF of the $(2\beta+1)$-dimensional vector $\boldsymbol{X}_{\text{joint}}=[X_{m-\beta}, X_{m-\beta+2}, \dots, X_m, \dots, X_{m+\beta-2}, X_{m+\beta}]^T$ follows a multivariate normal distribution:

$$f(\boldsymbol{X}_{\text{joint}}) = \frac{1}{(2\pi)^{\frac{2\beta+1}{2}}|\Sigma_{\text{joint}}|^{\frac{1}{2}}}\exp\left(-\frac{1}{2}\boldsymbol{X}_{\text{joint}}^T\Sigma_{\text{joint}}^{-1}\boldsymbol{X}_{\text{joint}}\right), \qquad (B2)$$

while the PDF of the $2\beta$-dimensional neighbor vector $\boldsymbol{X}_t$ is given by:

$$f(\boldsymbol{X}_t) = \frac{1}{(2\pi)^{\frac{2\beta}{2}}|\Sigma_t|^{\frac{1}{2}}}\exp\left(-\frac{1}{2}\boldsymbol{X}_t^T\Sigma_t^{-1}\boldsymbol{X}_t\right), \qquad (B3)$$

where $\Sigma_t$ is the submatrix derived from $\Sigma_{\text{joint}}$ y removing the $m$-th row and column. From the distributions, the conditional PDF of the target variable is defined as:

$$f(X_m \mid \boldsymbol{X}_t) = \frac{f(\boldsymbol{X}_{\text{joint}})}{f(\boldsymbol{X}_t)}. \qquad (B4)$$

For multivariate Gaussian variables, this conditional PDF $f(X_m \mid \boldsymbol{X}_t)$ remains a Gaussian distribution with a conditional mean $\mu_{m|t}$ and a conditional variance $\sigma_{m|t}^2$. Building upon this, the discrete conditional probability and the resulting conditional min-entropy are derived to yield the final analytical expressions provided in Eq.7 and Eq. 8, which are not reiterated here for brevity.

**Appendix C. conditional min-entropy under electric noise**

Owing to their distinct physical origins, the quantum phase noise $X$ and the electronic noise $E$ are regarded as statistically independent. Consequently, as indicated by Eq. 3, the measured raw data $Y$ remains a multivariate Gaussian distribution. To quantify the uncertainty, we first establish the conditional PDF constrained by the side information $E$. Based on the properties of multivariate normal distributions, the joint PDF given a specific noise instance $e$ is expressed as:

$$f\left(\boldsymbol{Y}_{\text{joint}}\middle|E=e\right)=\frac{1}{(2\pi)^{\frac{2\beta+1}{2}}|\Sigma_{Y\text{joint}}|^{\frac{1}{2}}}\cdot\exp\left(-\frac{1}{2}(\boldsymbol{Y}_{\text{joint}}-\boldsymbol{e}_{\text{joint}})^{T}\Sigma_{Y\text{joint}}^{-1}(\boldsymbol{Y}_{\text{joint}}-\boldsymbol{e}_{\text{joint}})\right).\quad (C1)$$

Similarly, the marginal PDF of the neighboring variables $\boldsymbol{Y}_t$, denoted $f(\boldsymbol{Y}_t|E=e)$, is obtained by marginalizing the target variable $Y_m$ from the joint PDF. This also results in a multivariate Gaussian distribution:

$$f(\boldsymbol{Y}_t|E=e)=\frac{1}{(2\pi)^{\frac{2\beta}{2}}|\Sigma_{Yt}|^{\frac{1}{2}}}\cdot\exp\left(-\frac{1}{2}(\boldsymbol{Y}_t-\boldsymbol{e}_t)^{T}\Sigma_{Yt}^{-1}(\boldsymbol{Y}_t-\boldsymbol{e}_t)\right),\quad (C2)$$

where the covariance matrix $\Sigma_{Yt}$ is the submatrix derived from $\Sigma_{Y\text{joint}}$ by removing the row and column associated with $Y_m$.

Following the same calculation process as **Accounting for correlation**, we can obtain:

$$\begin{aligned}P(Y_m^{\text{dis}}=y_m \mid \boldsymbol{Y}_t^{\text{dis}}=\boldsymbol{y}_t,E)&=\frac{P\left(Y_m^{\text{dis}}=\text{y},\boldsymbol{Y}_t^{\text{dis}}=\boldsymbol{y}_t\right)|E}{P\left(\boldsymbol{Y}_t^{\text{dis}}=\boldsymbol{y}_t\right)|E}\\&=\frac{\int_{D_{Ym}}\int_{D_{Yt}}f(\boldsymbol{Y}_{\text{joint}}|E=e)\,dY_m\,d\boldsymbol{Y}_t}{\int_{D_{Yt}}f\,(\boldsymbol{Y}_t|E=e)\,d\boldsymbol{Y}_t}.\end{aligned}\quad (C3)$$

Therefore, the final expression for the conditional min-entropy in the presence of electrical noise is given by:

$$\begin{aligned}&H_{\min}(Y_m^{\text{dis}} \mid \boldsymbol{Y}_t^{\text{dis}},E)\\&=-\log_2\left(\max_{y_m,\boldsymbol{y}_t}P\left(Y_m^{\text{dis}}=y_m \mid \boldsymbol{Y}_t^{\text{dis}}=\boldsymbol{y}_t,E\right)\right).\end{aligned}\quad (C4)$$

Additionally, by adopting a simplification strategy consistent with our analysis of the ideal scenario, we utilize the dominant correlation between adjacent samples to reduce the multi-dimensional evaluation into a bivariate framework, which leads to an explicit analytical expression for the conditional min-entropy, parameterized by the electrical noise and the SNR. The complete mathematical derivation is detailed in **Practical consideration**.